\documentclass{chjaa}
\usepackage{graphicx,times}
\input{epsf.sty}
\input{psfig.sty}
\begin{document}

   \title{Formation of Transient Coronal Holes during Eruption of a Quiescent
   Filament and its Overlying Sigmoid\,$^*$
   \footnotetext{$*$ Supported by the National Natural Science Foundation of China.}
}

   \volnopage{Vol.\ 8 (2008), No. 3,~ 000--000}
   \setcounter{page}{1}

   \author{Li-Heng Yang
      \inst{1,2}
   \and Yun-Chun Jiang
      \inst{1}
   \and Dong-Bai Ren
      \inst{1,2}
          }

   \institute{National Astronomical Observatories/Yunnan Observatory, Chinese Academy of Sciences, Kunming
   650011; \emph{yangliheng@ynao.ac.cn}
         \and
             Graduate school of Chinese Academy of Sciences,
             Beijing 100049}

   \date{Received~~2007 May 15; accepted~~2008 February 4}

   \abstract{
By using H$\alpha$, He I 10830, EUV and soft X-ray (SXR) data, we
examined a filament eruption that occurred on a quiet-sun region
near the center of the solar disk on 2006 January 12, which
disturbed a sigmoid overlying the filament channel observed by the
$\emph{GOES-12}$ SXR Imager (SXI), and led to the eruption of the
sigmoid. The event was associated with a partial halo coronal mass
ejection (CME) observed by the Large Angle and Spectrometric
Coronagraphs (LASCO) on board the Solar and Heliospheric Observatory
($\emph{SOHO}$), and resulted in the formation of two flare-like
ribbons, post-eruption coronal loops, and two transient coronal
holes (TCHs), but there were no significantly recorded $\emph{GOES}$
or H$\alpha$ flares corresponding to the eruption. The two TCHs were
dominated by opposite magnetic polarities and were located on the
two ends of the eruptive sigmoid. They showed similar locations and
shapes in He I 10830, EUV and SXR observations. During the early
eruption phase, brightenings first appeared on the locations of the
two subsequent TCHs, which could be clearly identified on He I
10830, EUV and SXR images. This eruption event could be explained by
the magnetic flux rope model, and the two TCHs were likely to be the
feet of the flux rope.
   }

   \authorrunning{L. H. Yang, Y. C. Jiang \& D. B. Ren }
   \titlerunning{Filament Eruption, Transient Coronal Holes and CME}

   \maketitle

\section{Introduction}
\label{sect:intro}

Coronal mass ejections (CMEs) are sudden eruptions of magnetized
plasma from the solar corona into the heliosphere that can represent
a large-scale rearrangement of the coronal magnetic field. CMEs have
been considered to be the cause of interplanetary shocks and the
driver of geomagnetic storms. In particular, halo CMEs have a great
influence on space weather, because they may be directed earthward
and so can lead to significant geo-disturbances (jing et
al.~\cite{jin04}). Although the halo CMEs have been extensively
investigated, their origin and initiation are still not well
understood. Therefore, identifying the on-disk CME signatures is
vital for detecting earth-directed CMEs and for an understanding of
the ultimate driving mechanism of CMEs, as well as for the forecast
of space weather. Various warning signs of CMEs have been identified
by in recent work, such as flares, eruptions of filaments and
sigmoidal structures, transient coronal holes (TCHs), and so on
(Hudson \& Cliver~\cite{hud01}). However, the relationship between
the on-disk surface activities and CMEs needs to be further checked.
In a few CME events, very weak surface activities led to an
earthward CME. For example, the event on 1997 January 6 showed that
a very weak and unimpressive solar activity induced a halo CME and a
geomagnetic storm (Webb~\cite{web98}). Another similar event
reported by jiang et al. (\cite{jia06}) showed that a filament
eruption without associated H$\alpha$ and $\emph{GOES}$ flare was
related to a partial halo CME. These weak surface activities are
significant to predict the space weather, and so cannot be ignored.

TCHs have been identified as key on-disk indicators of CMEs in some
halo events with observations from $\emph{Yohkoh}$ Soft X-ray (SXR)
Telescope and the Extreme Ultraviolet Imaging Telescope (EIT) on
board the $\emph{Solar and Heliospheric Observatory}$
($\emph{SOHO}$) (Thompson et al.~\cite{tho98}; Wang et
al.~\cite{wan00}). It has been found that they are often associated
with eruptions of filaments or sigmoidal structures (Sterling et
al.~\cite{ste00}; Jiang et al.~\cite{jia07b}), and appear as
darkenings in SXR and EUV and brightenings in He I 10830\,{\AA}
(Toma et al.~\cite{tom05}) in regions of unipolar opening magnetic
field, with lifetimes of about a few hours to days generally, and
can give rise to transient high-speed wind (Rust~\cite{rus83}).

Since their typical timescale of formation is less than an hour,
much shorter than the typical radiative cooling timescale of about
36 hours in the corona (Hudson et al.~\cite{hud96}), the TCHs are
often interpreted as due to density depletions, rather than
temperature variation. A pair of compact and symmetric TCHs
dominated by opposite polarities can form immediately close to the
two ends of an eruptive filament or, sometimes, a sigmoid. It is
believed that they mark the positions of the footpoints of the flux
rope and the mass loss from them is expelled into the CMEs (Sterling
\& Hudson~\cite{ste97}; Webb et al.~\cite{web00}; jiang et
al.~\cite{jia06}).

Moreover, such dual TCHs can have H$\alpha$ counterparts, which
means that they could extend from the corona deep into the
chromosphere (Jiang et al.~\cite{jia03}). More recently, Toma et
al.(\cite{tom05}) and Jiang et al.(\cite{jia07a}) found that some
twin TCHs showed up above a clear chromospheric network pattern
consisting of patchy bright/dark H$\alpha$/He I 10830\,{\AA} plages
with strongly concentrated magnetic network elements, as
darkenings/brightenings of these plages during eruptions of the
associated filament. It was further found that these TCHs were often
preceded by chromospheric and coronal brightenings in the rising
phase of the associated flares. Up to now, however, only a few such
cases were observed, and the cause and role of the plages and
preceding brightenings in the formation of TCHs are still not clear.
To understand clearly the relationship of the brightenings and the
following TCHs, and the eruption process and the associated CMEs,
further observations are strongly needed.

In this paper, we present multi-wavelength observations of the
eruption of an S-shaped filament that occurred on 2006 January 12,
in which twin TCHs were formed in He I 10830\,{\AA}, EUV, and SXR.
This eruption was not associated with any recorded $\emph{GOES}$ or
H$\alpha$ flare but was directly related to a partial halo CME
observed by the Large Angle and Spectrometic Coronagraphs (LASCO) on
board $\emph{SOHO}$. We will show that the brightenings clearly
appeared at the locations of the following TCHs and the TCHs were
possibly part of the on-disk proxy of the CME source region.

\section{Observations}
\label{sect:Obs} For the present work, the following data are used:

1 Full-disk He I 10830 {\AA} intensity and velocity images from the
Chromospheric Helium Imaging Photometer (CHIP, MacQueen et
al.~\cite{mac98}) and H$\alpha$ disk images from the Polarimeter for
Inner Coronal Studies (PICS) at the Mauna Loa Solar Observatory
(MLSO). The CHIP data were acquired using a tunable Lyot filter
($\approx$0.13 bandpass) positioned at seven wavelengths covering
the spectral region from 10826 to 10834\,{\AA}, which provides a
measure of the line-of-sight velocity of a filament over
100\,km~$s^{-1}$. The cadence of these images is 3 minutes and the
pixel size is $2.3''$.  The PICS disk images were acquired by using
a narrowband filter of $\pm$0.5 {\AA} which allows for line-of-sight
velocities about 40--45\,km~$s^{-1}$. These images have the same
cadence with the CHIP data and a resolution of $2.9''$\,pixel$^{-1}$
(Toma et al.~\cite{tom05}).

2. Full-disk EUV images from the EIT
(Delaboudini$\acute{e}$re~\cite{del95}) on $\emph{SOHO}$. EIT images
are taken in four spectral bands centered on Fe {\sc ix/x}
(171\,{\AA}), Fe {\sc xii} (195\,{ \AA}), Fe {\sc xv} (284\,{\AA}),
He {\sc ii} (304\,{\AA}), which allow imaging of the solar plasma at
temperatures ranging from 6 ${\times}$10$^{4}$\,K to 3
${\times}$10$^{6}$\,K. For the present work, the EIT provided
195\,{\AA} images with cadence of 12 minutes and pixel size of
$2.6''$, while the 304\,{\AA} and 284\,{\AA} images were taken only
once every 6 hours.

3. Full-disk SXR images from the $\emph{GOES-12}$ Solar X-ray Imager
(SXI, Hill et al.~\cite{hil05}). For the current work, SXR images
were taken every two minutes using thin polyimide filter. These
images cover a wavelength range of 0.6-60\,nm (sensitive to the
temperature of 0.9-20\,MK), at a pixel size of $5''$.

4. Full-disk longitudinal magnetograms from the Michelson Doppler
Imager (MDI, Scherrer et al.~\cite{sch95}) on $\emph{SOHO}$, with
cadence of 96 minutes and pixel size about $2''$.

5. C2 and C3 white-light coronagraph data from LASCO,
which cover the range of 2 to 6 and 4 to 32 solar
radii, respectively (Brueckner et al.~\cite{bru95}).

\section{results}
\label{sect:res}

\begin{figure}
   \begin{center}
   \mbox{\epsfxsize=0.9\textwidth\epsfysize=0.9\textwidth\epsfbox{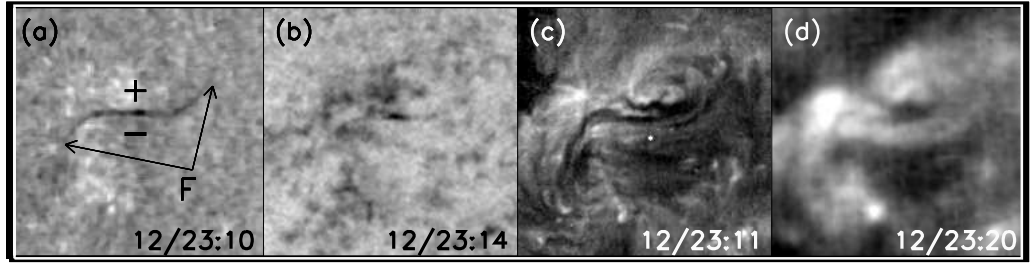}}
   \caption{\baselineskip 3.6mm
   MLSO/PICS H$\alpha$ line center (a), MLSO/CHIP
   He I 10830\,{\AA} intensity (b),
   EIT 195\,{\AA} (c), and SXI SXR (d), showing the appearance of the
   filament, ``F", before its eruption.
   The field of view (FOV) is $475''$${\times}$$475''$.}
   \label{Fig:plot1}
   \end{center}
\end{figure}

\begin{figure}
   \begin{center}
   \mbox{\epsfxsize=0.9\textwidth\epsfysize=0.9
   \textwidth\epsfbox{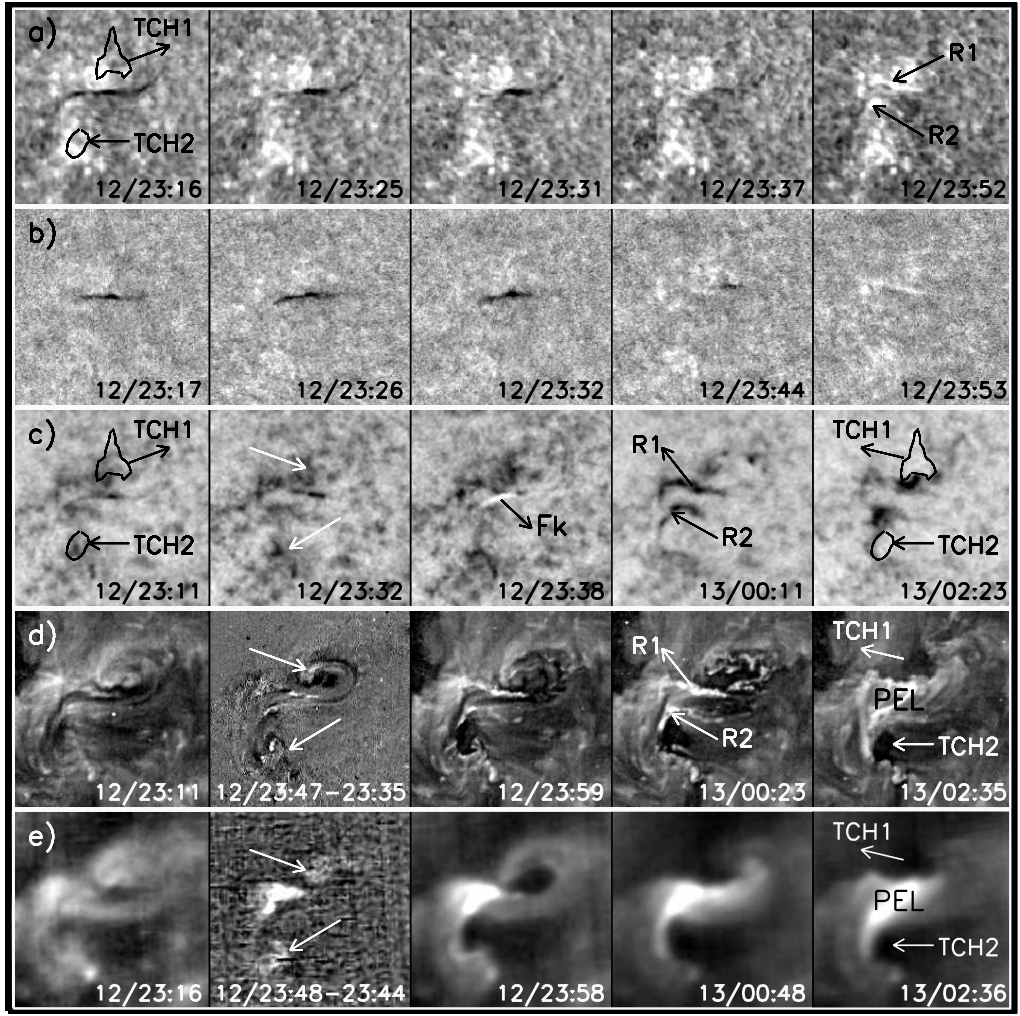}}
   \caption{\baselineskip 3.6mm
   MLSO/PICS H$\alpha$ line center (a), MLSO/CHIP
   He I 10830\,{\AA} velocity (b) and
   intensity (c),
   EIT 195\,{\AA} (d) and  SXI SXR (e), showing the evolution of the event.
   To clearly exhibit the brightenings in the early formation phase of the TCHs, a
   difference image is presented in the second column of (d)/(e).
   The FOV is the same as in Fig.~1.}
   \label{Fig:plot2}
   \end{center}
\end{figure}

\begin{figure}
   \begin{center}
   \mbox{\epsfxsize=0.8\textwidth\epsfysize=0.8\textwidth\epsfbox{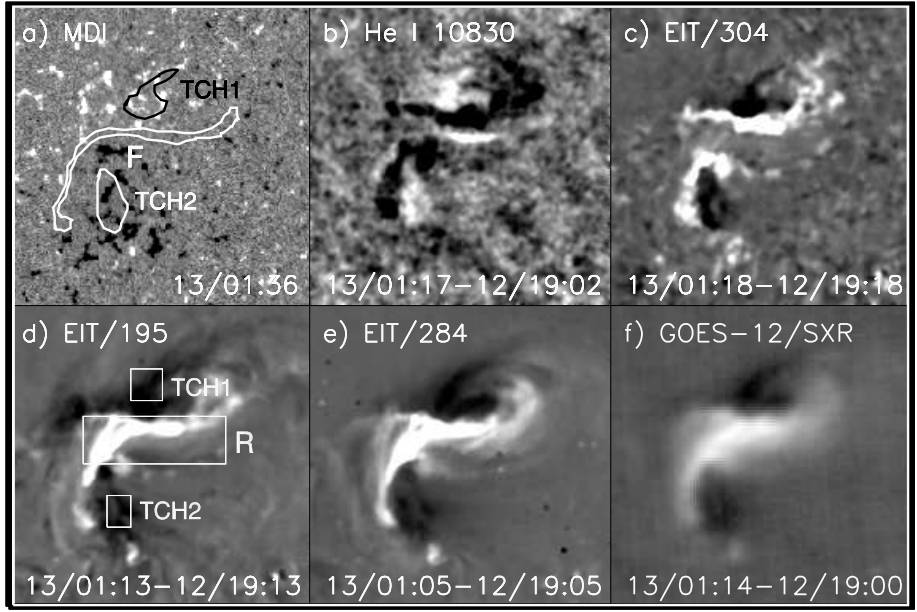}}
   \caption{\baselineskip 3.6mm MDI magnetogram (a), MLSO/CHIP He I 10830\,{\AA} (b),
   EIT 304 (c), 195 (d), and 284\,{\AA} (e), and SXI/SXR (f) difference images.
   The two TCHs, ``TCH1" and ``TCH2" are predominated by opposite polarities
   and are located on the two F ends. The TCHs and the pre-eruptive H$\alpha$ F are
   depicted by the black and white contours of (a).
    The FOV is the same as in Fig.~1. }
   \label{Fig:plot3}
   \end{center}
\end{figure}

\begin{figure}
   \begin{center}
   \mbox{\epsfxsize=1.0\textwidth\epsfysize=1.0\textwidth\epsfbox{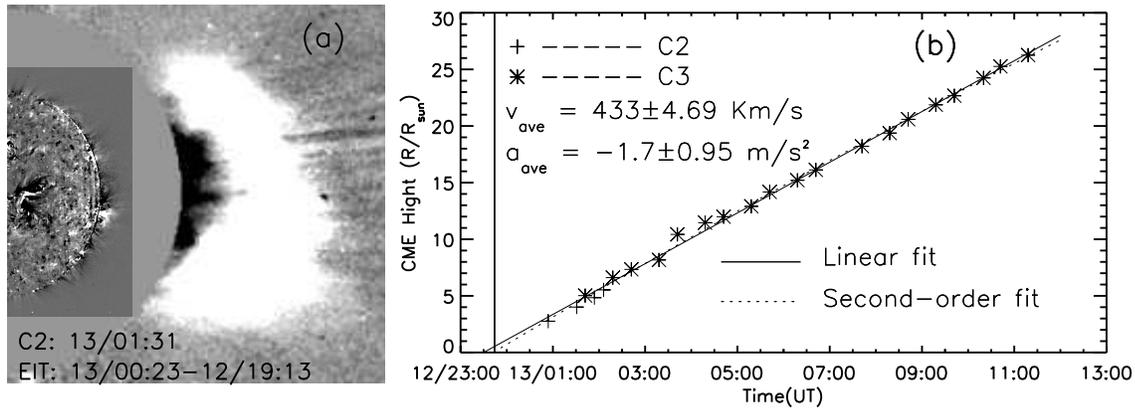}}
   \caption{\baselineskip 3.6mm (a) Composite image of the inner EIT 195\,{\AA} and the outer
    LASCO C2 difference images. (b) Height of the CME front as the function of time.}
   \label{Fig:plot4}
   \end{center}
\end{figure}

\begin{figure}
   \begin{center}
   \mbox{\epsfxsize=0.7\textwidth\epsfysize=0.7\textwidth\epsfbox{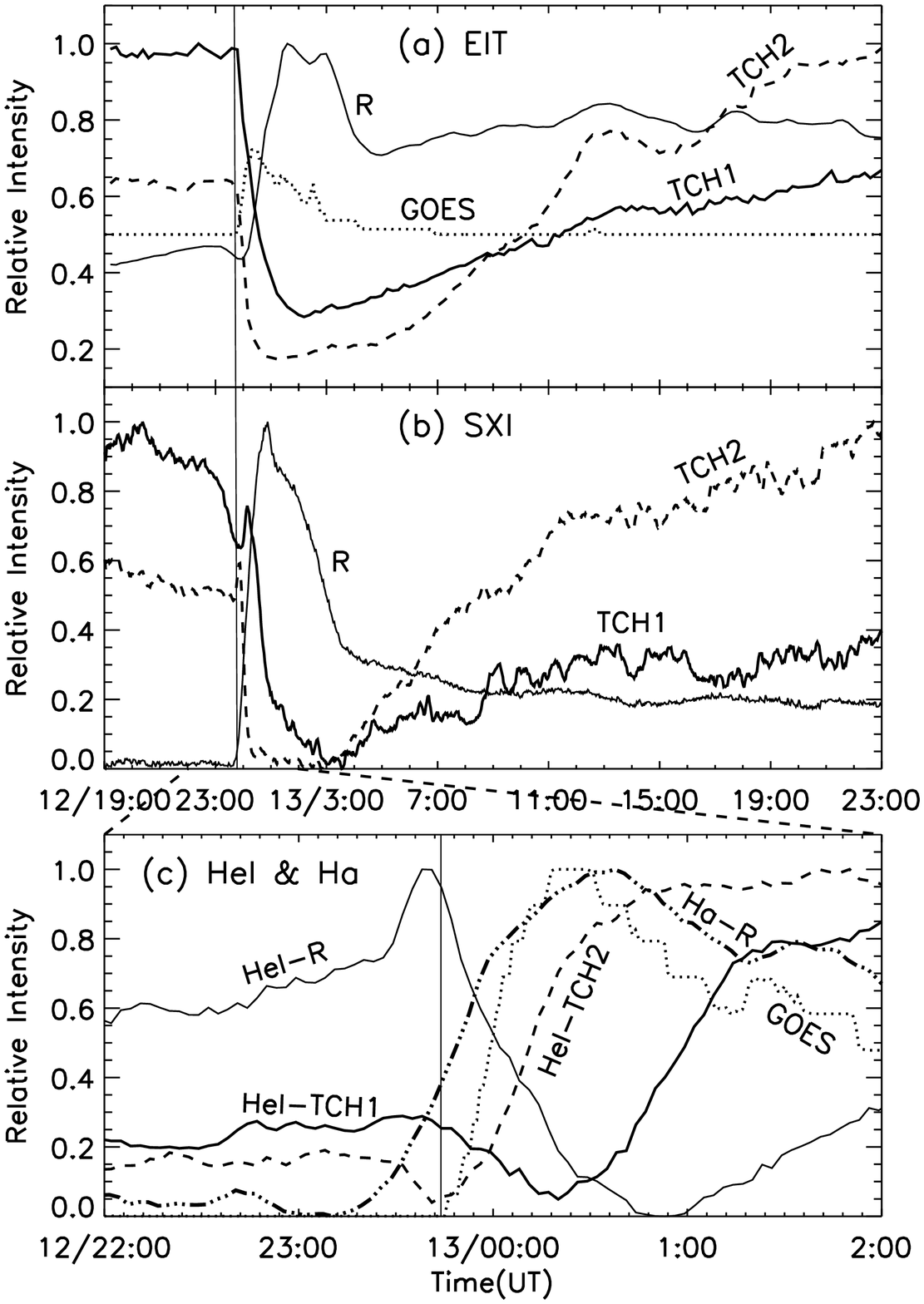}}
   \caption{\baselineskip 3.6mm Time profiles of $\emph{GOES-12}$/SXR in the energy channel of 1-8
   {\AA} (dashed thin lines in Panels a and c), in arbitrary
   unites and in unites normalized to one, respectively. The light curves of EIT 195\,{\AA} (a),
    SXI/SXR (b), He I 10830\,{\AA} (c) intensities in
    areas centered on the TCH1, TCH2, and R(indicated by the boxes in Fig.~3(d)),
    and H$\alpha$ intensities (c) in an area centered on the R are
   computed from the intensity averaged over these regions. EIT 195\,{\AA} is normalized
   to the maximum value, and the SXI/SXR, He I 10830 and H$\alpha$
   are normalized to one. }
   \label{Fig:plot5}
   \end{center}
\end{figure}

The eruptive filament, ``F", was located on a quiet-sun region in
the southern hemisphere centered at S03W16, on January 12. Figure~1
shows the general appearance of the eruptive region before the F
eruption in the H$\alpha$, He I 10830\,{\AA}, EIT 195\,{\AA}, and
SXR images, respectively. In the H$\alpha$ image (Fig.~1a), F,
indicated by the two black arrows, clearly shows an inverted S
shape. As expected, it lies along a polarity inversion zone of the
photospheric magnetic fields, and the corresponding
positive/negative polarity on either side is marked by plus/minus
sign. When F can be identified as a dark feature in He I
10830\,{\AA} (Fig. 1b), its EUV counterpart was clearly seen and
showed a similar inverted S shape in the EIT 195\,{\AA} image (Fig.
1c). However, F in EUV appeared more complicated and bifurcated, and
the same bifurcation was seen in the SXR image. In the higher
temperature wavebands, F appeared increasingly broader. From SXI/SXR
image (Fig.~1d), we also note an inverted S sigmoid located just
above the F, indicating that they could belong to the same
topological structure. This is consistent with the observation of
Pevtsov (\cite{pev02}) that the chromospheric filament and the
coronal sigmoid had a close spatial association. Although we can not
determine F's chirality from its barb orientations due to the poor
spatial resolution of the $H{\alpha}$ observations, we conclude that
the axial field of F before its eruption was directed eastward since
the sign of the photospheric magnetic field in the F ends was
negative in the southeast and positive in the northwest (see Fig.
3a). According to the definition given by Martin et al.
(\cite{mar94}), F was dextral. However, it is noted that the
 chirality pattern consistent with both the dextral F and the inverted S
sigmoid disobeyed the hemispheric chirality rule for Fs and sigmoids
in the southern hemisphere (Zirker et al.~\cite{zir97}).

Figure 2 shows the evolution of the eruption in H$\alpha$, He I
10830\,{\AA}, EIT 195\,{\AA}, and SXI/SXR. The eruption showed up as
a total disappearance of F in H$\alpha$. We see that its inverted S
shape was clearly discernible at 23:16\,UT, but parts of it became
invisible at 23:25\,UT, and the whole F disappeared at 23:37\,UT. In
the He I 10830\,{\AA} velocity observations, however, F mainly
showed blueshift signature, clearly indicating that the
disappearance was due to its eruption. We also note that the
blueshift signature could be detected at 23:44\,UT, lagging the
complete disappearance of F in H$\alpha$ by 7 minutes. This possibly
suggests that the erupting F was fast enough to be Doppler shifted
out of the filter of the MLSO/PICS instrument but was still in the
detectable range of the MLSO/CHIP instrument. In EIT 195 \AA\
images, we see that the dark EUV counterpart of the H$\alpha$ F also
showed a consistent eruptive process. As a distinct feature of the
event, however, the SXR sigmoid was strongly disturbed during the
eruption, and it appeared that the sigmoid also underwent an
eruption, which was made manifest by the clear disappearance of some
loops that contained it after the F eruption. Thus, this event
involved the eruptions of both the F and the sigmoid. This is
greatly different from the cases studied by Pevtsov (\cite{pev02})
and jiang et al. (\cite{jia07b}), in which coronal sigmoids
underwent activations and eruptions while the underlying H$\alpha$
filaments were left largely untouched. The F eruption was
accompanied by the occurrence of two flare-like ribbons, ``R1" and
``R2", in H$\alpha$, He I 10830 {\AA} and EIT 195 {\AA} lines. They
were located on opposite sides of the eruptive F, and were dark in
He I 10830\,{\AA} due to the increased absorption. Notably, a bright
flare-like kernel, ``Fk", was visible from 23:35 to 23:44\,UT in He
I 10830 {\AA} intensity images, this means that the energy release
was large enough to change the He I 10830 {\AA} absorption line into
emission (Penn \& Kuhn~\cite{pen95}).

After the F eruption, post-eruptive loops, ``PELs", gradually
appeared in EUV and SXR observations. They connected the two
flare-like ribbons, and their footprints expanded with the
increasing separation of the ribbons. All of these observations
indicated typical features of flares associated with F eruptions,
but no optical flare was reported by the Solar Geophysical Data
online around the time of the event. Furthermore, despite the
$\emph{GOES}$ SXR flux showing an increase relative to the
background level after the F eruption (see Fig.~5), no $\emph{GOES}$
flare above X-ray class B1 was recorded. Therefore, it seems that
the flare-like ribbons were too weak to be regarded as an H$\alpha$
or $\emph{GOES}$ flare. As the SXR and H$\alpha$ wavebands each
contain only 5--10\% of the total radiated energy, it may not be
proper to identify a flare with only these wavebands (Zhou et
al.~\cite{zho03}). The EIT 195\,\AA\ light curve measured in the
flare-like region is plotted in the Figure~5(a), and we note that
the brightness increased about 125\% from the starting time to the
time of maximum enhancement. Due to the brightness at the start time
was greater than that of the background, the enhancement could be
considered as an EUV flare according to the criterion given by Zhou
et al. (\cite{zho03}).

As another remarkable feature of the event, two TCHs, labelled
``TCH1" and ``TCH2", were formed during the eruptions of the F and
the sigmoid. After the eruptions, they were quite obvious in He I
10830\,{\AA} intensity, EIT 195\,{\AA}, and SXI/SXR images, see
Figure~2. As in the cases investigated by jiang et al.
(\cite{jia03}), Toma et al. (\cite{tom05}) and jiang et al.
(\cite{jia07a}), it is noted that the dual TCHs were also preceded
by faint brightenings in their early formation phase, which can be
clearly seen in the He I 10830\,{\AA} intensity image, and the EVU
and SXR difference images; see the second row of Figure~2 (indicated
by the white arrows). They were located at the sites of the
following TCHs, and were kept away from the two flare-like ribbons,
so were not the result of a spreading and expansion of these
ribbons. Thus it appears that formation of TCHs is often associated
with certain brightenings. Figure~3 shows the difference images from
the pre-event images at He I 10830\,{\AA}, 304\,{\AA}, 195\,{\AA},
284\,{\AA} and SXI/SXR lines. The TCHs, appearing as brightening
regions in He I 10830\,{\AA} and dimming regions in EUV and SXR, had
similar locations and shapes in these lines formed over a
temperature range from several $10^{4}$ K to 2${\times}$10$^{6}$ K,
so indicating that they are caused by density depletions rather than
by temperature variations (Thompson et al.~\cite{tho98}). Moreover,
they occurred in opposite polarity regions near the two ends of the
erupted sigmoid and flanked the central PEL.

A partial halo CME was observed by SOHO/LASCO around the time of the
event. The CME was first detected by LASCO C2 at 00:54\,UT on
January 13, and later became visible in C3 at 02:42\,UT located
above the W limb. According to the measurements of Seiji Yashiro,
its center position angle (PA) was $281\dg$ with a width of
$126\dg$. Figure~4a presents a composite image of an inner EIT
195\,{\AA} with an outer LASCO C2 difference image. We see that the
eruptive region was nearly located along the direction of the
central PA of the CME. The height-time (H-T) plot of the CME front
at $267\dg$ is shown in Figure~4b. The average speed of the CME
front given by a linear fitting was 433$\pm$4.96~km $s^{-1}$, and
the average acceleration given by a second-order polynomial fitting
was -1.7$\pm$0.95 m~$s^{-2}$ (shown in the upper left corner of
Fig.~4b), indicating that the CME was a slow one, which is
consistent with the results given by sheeley et al. (\cite{she99})
that slow CMEs were often associated with filament eruptions. From
the back extrapolation of the second-order polynomial fitting of the
CME H-T plot to the solar disk center, we estimate that the onset
time of the CME was about 23:45\,UT on January 12 (marked by a
vertical bar in Fig.~4 (b)), which was very close to the start time
of the $\emph{GOES}$ 1-8 {\AA} SXR flux increase and the onset time
of the brightenings preceding the appearance of the two TCHs(marked
by vertical bars in Fig.~5). These consistent spatial and temporal
behaviors indicated that the F eruption and the two TCHs were
directly related to the CME.

Figure 5 presents the profile of $\emph{GOES}$ 1-8\,{\AA} SXR flux
and the light curves of EIT 195\,{\AA}, SXI/SXR, and He I
10830\,{\AA} intensities of the two TCHs and the flare-like regions
and H$\alpha$ intensities of the flare-like region, respectively.
Beginning at 23:45 UT on January 12 (marked by the vertical line in
Fig.~5), the $\emph{GOES}$ SXR flux showed a weak increase, taking
about 1 hour to reach the peak, which was below the X-ray class B1
level. So the enhancement was too weak to be recorded as a
$\emph{GOES}$ flare. Furthermore, we see that EIT 195\,{\AA},
SXI/SXR light curves measured in the flare-like region are similar
to the $\emph{GOES}$ 1-8\,{\AA} SXR flux profile, and the
enhancement of the flare-like region began at about 23:45\,UT,
January 12. However, the start time of enhancement of the flare-like
region in He I 10830 and H$\alpha$ light curves was earlier the time
of disappearance of F. Especially, with the appearance of the
flare-like kernel, the light curve showed an increase between 23:35
to 23:44\,UT, January 12. We also note that during the formation of
the two TCHs, the EIT 195\,{\AA} and SXR intensities obviously
decreased and the He I 10830\,{\AA} intensity distinctly increased.
In addition, we found that during the early eruption phase there was
an increase in the SXR flux curves and a decrease in the He I
10830\,{\AA} flux curves due to the appearance of the brightenings,
and the increase (decrease) lasted only during the rise phase of
$\emph{GOES}$ 1-8 {\AA} SXR flux curve.

\section{Conclusions and Discussion}
\label{sect:con}

We have made an investigation on a filament eruption near the center
of the solar disk on 2006 January 12, and on the associated
brightenings, TCHs, and a halo CME, as well as the eruptive sigmoid
above the filament. The main observational results are as follows:
(1) The disappearance of the inverted S-shaped filament was followed
by two flare-like ribbons, two TCHs, and post-eruption coronal
loops, but no evident X-ray and H$\alpha$ flares were recorded. (2)
During the early eruption phase, brightenings first appeared on the
sites of the two TCHs, and were short lived during the rise phase of
the flare, then the two TCHs were formed near the two ends of the
sigmoid, dominated by opposite polarities, which were clearly
visible in He I 10830\,{\AA}, EUV and SXR, with similar locations
and shapes. (3) The partial halo CME showed a close spatial and
temporal relation to the filament eruption and the two TCHs.

This event was not related to any recorded H$\alpha$ and
$\emph{GOES}$ flares, while the usual feature of flares of two
flaring ribbons, was identified in multi-wavelength observations
(H$\alpha$, He I 10830\,{\AA}, EIT 195\,{\AA}), but it is hard to
associate such a weak surface activity to the partial halo CME.
Fortunately, the filament and the sigmiod eruption, and the
formation of the two TCHs, which are considered to be predictors of
the CME, directed our attention to this event. It is incredible that
such a very minor change in the chromosphere could bring about such
a major coronal perturbation. Similar results have been obtained by
Webb et al. (\cite{web98}), and they found that some weak on-disk
activities on 1997 January 6 led to a CME and the ``problem''
geomagnetic storms. Later, Shakhovskaya et al. (\cite{sha02}) showed
that a prominence eruption on 2000 August 11 with no associated
flares resulted in a faint CME. More recently, jiang et al.
(\cite{jia06}) reported a filament eruption on 1999 March 21 without
flares recorded, which led to a partial halo CME. These events
indicate that CMEs may or may not be relate to flares in the
chromosphere ($\check{S}$vestka~\cite{sve01}) and that weak events
are valuable in identifying the source regions of the earth-directed
CMEs, important for predicting space weather.

We have described a rare event where the eruptions of a quiescent
filament and an overlying sigmoid led to the formation of the two
TCHs, which is not related to any recorded flare events, which is
very similar to the event of 1997 October 23 presented by Hudson and
Cliver (\cite{hud01}). The H$\alpha$ filament and sigmoid are
considered to be the signature of a flux rope system
(Pevtsov~\cite{pev02}), and the eruption event can be explained by
the flux model of CME, the two TCHs representing the two footprints
of the flux rope (Sterling and Hudson~\cite{ste97}; Jiang et
al.~\cite{jia03}; Jiang et al.~\cite{jia06}). However, during this
event both the filament and the sigmoid erupted, unlike the
observations by Pevtsov (\cite{pev02}) where the sigmoid activated
and disappeared without a corresponding filament eruption.

According to the analytic 3-D MHD model of flux ropes proposed by
Gibson and Low (\cite{gib00}), the two TCHs would appear
anti-symmetrically on either side of the photospheric neutral line,
and interpreted to be the counterpart of the CME cavity seen at the
limb. The formation of the two TCHs could be explained by the
opening of previous closed field lines or by the expansion of those
related to the flux rope. The two TCHs clearly appeared on the
304\,{\AA} difference image, indicating that the feet of the flux
rope were deeply rooted in the chromosphere, and that the mass loss
may originate from the cooler and denser chromosphere. Meanwhile,
the two TCHs had a temperature range from several $10^{4}$\,K to
2${\times}$10$^{6}$\,K, suggesting that the appearance of the two
TCHs was due to the density depletion rather than the temperature
decrease. The mass loss probably provided a supplement to the CME.
In addition, the brightenings appeared on the locations of the two
TCHs at the initial phase of the flare, and their appearance may be
the result of a flux rope eruption (jiang et al.~\cite{jia07a}).
This has not been described by any theoretical models, and further
observational and theoretical investigations are needed for further
understanding.

\begin{acknowledgements}

The authors thank an anonymous referee for valuable comments. The
H$\alpha$ and He I 10830\,{\AA} data are provided by the High
Altitude Observatory, which is part of the National Center for
Atmospheric Research under sponsorship of the National Science
Foundation. The authors are indebted to the $\emph{SOHO}$/EIT, MDI
and LASCO teams, $\emph{GOES}$/SXI team for free access to the
wonderful data. This work is supported by the NSFC under grants
10573033 and 40636031, and by the 973 program (2006CB806303).

\end{acknowledgements}

\label{lastpage}

\end{document}